\documentclass[11pt]{article}
\usepackage[T1]{fontenc}
\usepackage{kpfonts}
\usepackage[dvipsnames]{xcolor}
\usepackage{pdfpages}
\usepackage{sgame}
\usepackage{accents}
\usepackage{tikz-cd}
\usepackage{float}
\usepackage[round]{natbib}   
\usepackage{multirow}
\usepackage{dutchcal}
\usepackage{pdfpages}
\usepackage{bbm}
\usepackage{sgame}
\usepackage{mathtools}
\usepackage{amsthm}
\usepackage{enumitem}
\usepackage[margin=1in]{geometry}
\usepackage{caption}
\usepackage{subcaption}
\usepackage{epigraph}
\usepackage{listings}
\usetikzlibrary{calc}
\usetikzlibrary{shapes,arrows}

\DeclareMathOperator\supp{supp}

\newtheorem{theorem}{Theorem}[section]
\newtheorem{proposition}[theorem]{Proposition}
\newtheorem{lemma}[theorem]{Lemma}
\newtheorem{corollary}[theorem]{Corollary}

\theoremstyle{definition}

\newtheorem{remark}[theorem]{Remark}

\usepackage{hyperref}
\hypersetup{
    colorlinks=true,
    linkcolor=CadetBlue,
    filecolor=CadetBlue,      
    urlcolor=CadetBlue,
    citecolor=CadetBlue,
}

\definecolor{backcolour}{rgb}{0.63, 0.79, 0.95}

\lstdefinestyle{mystyle}{
  backgroundcolor=\color{backcolour},
  basicstyle=\ttfamily\footnotesize,
  breakatwhitespace=false,         
  breaklines=true,                 
  captionpos=b,                    
  keepspaces=true,                 
  numbers=left,                    
  numbersep=5pt,                  
  showspaces=false,                
  showstringspaces=false,
  showtabs=false,                  
  tabsize=2
}

\providecommand{\keywords}[1]{\textbf{\textit{Keywords:}} #1}
\providecommand{\jel}[1]{\textbf{\textit{JEL Classifications:}} #1}

\begin{document}
\author{Mark Whitmeyer\thanks{Arizona State University, Email: \href{mailto:mark.whitmeyer@gmail.com}{mark.whitmeyer@gmail.com} \newline I thank Vasudha Jain, Joseph Whitmeyer and Thomas Wiseman for their comments.} }

\title{Submission Fees in Risk-Taking Contests}

\date{\today}

\maketitle

\begin{abstract}
This paper investigates stochastic continuous time contests with a twist: the designer requires that contest participants incur some cost to \textit{submit} their entries. When the designer wishes to maximize the (expected) performance of the top performer, a strictly positive submission fee is optimal. When the designer wishes to maximize total (expected) performance, either the highest submission fee or the lowest submission fee is optimal.
\end{abstract}
\\
\jel{C72; C73; D81; D82; D83} 

\section{Introduction}

In many environments, agents compete for prizes by choosing what risks to take. Fund managers in a private equity firm choose how risky to make their portfolios, with the goal of outperforming their fellows for a bonus or promotion. Academics choose what subfields to inhabit and what projects to pursue so they can write successful grant proposals and publish in top journals. Graduate students in many fields prepare job market papers in order to secure one of the limited spots in academia. Filmmakers, TV showrunners, chefs and restaurateurs, and musicians decide how bold and adventurous to be in their work with the goal of winning a Palm d'Or, an Emmy, a James Beard award, or a ``Grammy.'' 

In many of these situations, agents are not candidates for a prize by default: each contestant must incur a cost to enter or declare her candidacy for a prize in some way. For instance, an investment firm may require a promotion aspirant to fill out a form to apply and insist that any candidate present evidence of her successes. Grants and awards in academia mandate that researchers put together proposals, the preparation of which takes a considerable amount of time and effort. Job market candidates spend scores of hours in the job market season preparing their applications. The major awards in film, TV, cooking, and music all require hopefuls to submit entries and pay various fees.\footnote{Moreover, note that the submission costs that we consider in this paper need not be actual physical, monetary, or time costs, but can be psychological. An agent may suffer disutility from losing and hence experience the ``fear of failure.''} 

In this paper, we follow \cite{seel2016continuous} (henceforth \hypertarget{ss}{SS}) and look at a continuous time, winner-take-all contest in which $n$ agents compete by deciding when to stop independent Brownian motions with drift that are absorbed at zero.\footnote{In fact, virtually all of our results can be established in the more general scenario in which agents decide when to stop non-negative time-homogeneous diffusion processes that are absorbed at $0$, but the specific form of the Brownian motion with drift provides structure that eases the discussion and exposition.} Importantly, the value of the agents' processes and stopping decisions are private information. In academia, each academic's research is done largely in private, and graduate students are largely unaware on the outcomes of other graduate students' research projects prior to the market. The composition of a fund manager's portfolio is kept close to the chest, and so each may know very little about what, or how well, her competitors are doing. The details of potential Academy Award winning films are not public until their releases, and many of the releases of these contestants are concurrent, during ``Oscar season.'' Furthermore, the usefulness of this approach of allowing agents to compete through randomization is well established in the literature.\footnote{\cite{fang} discuss and reference the variety of papers that follow this ``competition through randomization'' approach.} 

The innovation of this paper is the introduction of a submission cost into this environment. That is, now each contestant must pay a fee in order to submit her entry for the prize. At first glance, these costs seem totally wasteful--especially given our assumption that levying such tolls is not a meaningful sort of revenue for the principal, i.e., that submission fees correspond to a pure destruction of surplus. However, submission fees affect the equilibrium play in the contest and any positive changes in this behavior could potentially outweigh the direct losses due to the costs. 

We take the perspective of a principal and look at two possible objectives: maximization of total (expected) output and maximization of the (expected) performance of the top agent. Naturally, the principal could either be a literal principal, e.g., the head of a fund overseeing the managers; or society, which benefits from better allocation of capital, innovations, and research. Perhaps surprisingly, even though submission costs do not factor into the principal's utility (they are a negligible portion of her revenue), they may nonetheless benefit the principal. More specifically, when the drift is positive the total (expected) output is strictly increasing in the size of the submission fee. When the drift is negative this relationship is flipped and lower submission fees are better. Regardless of the sign of the drift, the (expected) maximal performance is increasing in the submission fee provided the submission fee is sufficiently small, and the maximal submission fee remains optimal for a positive drift. 

Here is what is happening. Because each agent's behavior only affects other agents (and the principal) through the value at which she stops and enters into the contest, we need only consider the game in which agents compete by choosing distributions over these values. In the unique symmetric equilibrium, each agent's strategy yields a continuous distribution over stopped values that is supported on some interval $\left[0,\bar{x}\right]$ and places an atom on $0$. An agent does not enter the contest if and only if she obtains value $0$. This structure follows from standard arguments: the new wrinkle is that the submission fee requires the addition of the mass point. Importantly, the other effect of the submission fee is to increase the support (lengthen the right tail) of each agent's equilibrium distribution. To see why this is so, it is useful to think of the driftless case, in which each agent's process is a martingale. As the mass point on $0$ increases in size, there is less and less measure to be spread over positive values, yet the expectation of the distribution must be the same. Thus, the upper bound of the support must increase. Given this it is easy to see why the results follow--longer experimentation (a higher upper bound) is good when the drift is positive or when the principal cares about the expectation of the maximal output and bad when the drift is negative.

It is important to keep in mind that this submission fee, which the contestant chooses whether to pay \textit{after} the outcome of her effort (randomization), is completely different from an entry fee, which the contestant must pay prior to participation (\textit{before} the outcome of her effort). Entry fees, in contests with endogenous entry,\footnote{See, e.g., \cite{fu2015contests}, \cite{liu2019optimal}, and \cite{ginzburg2021optimal}, all of whom look at contests with endogenous entry and the effects of entry costs and prize structures on agents' participation decisions and subsequent effort levels.} can benefit a principal by \textit{reducing} the intensity of competition, thereby inducing greater effort by the contestants. In contrast, the submission fees of this paper \textit{increase} the intensity of competition.

Related to this phenomenon are the findings by \cite{nutz2021reward} (henceforth \hypertarget{nz}{NZ}). They look at the effects of changing the prize structure in the model of \hyperlink{ss}{SS}, and find that more inegalitarian prize schedules lead to longer tailed distributions, which thus benefit the principal when the processes' drifts are positive or when she cares only about the expected output of the winner. The contribution of this paper, therefore, is not to show that more intense competition in these contests with randomization is beneficial in certain circumstances, but to establish that submission fees are an instrument for ``turning up the heat.'' Moreover, our focus on the winner-take-all contest allows our discoveries to supplement those of \hyperlink{nz}{NZ}: we impose the most inegalitarian prize schedule, which is optimal when risk-taking is desired. As we discover, the principal can do even better by supplementing this structure with a submission fee.

\subsection{Related Work}

Several papers extend the results in \hyperlink{ss}{SS} and/or explore related questions, in a variety of ways. These papers include the already referenced work \hyperlink{nz}{NZ}, who look at optimal reward design in this setting; \cite{seel2015gambling}, who allows for heterogeneous loss constraints; \cite{nutz2021mean}, who look at a mean-field version of the game; and a trio of works, \cite{feng2015gambling}, \cite{feng2016gambling}, and \cite{feng2016gambling2}, who allow for more general diffusion processes, regret, and a random initial law, respectively. Special mention is due to \cite{fang}, who establish an equivalence between the stochastic contest of \hyperlink{ss}{SS} and a static game in which the contests choose randomizations over output that must satisfy a constraint on their mean (which game is itself a generalization of \cite{wagman2012choosing}). This finding leads directly to our equilibrium characterization and uniqueness result. They also look at the effects of different (more or less equal) prize schedules, contestant heterogeneity and incomplete information, and other contest modifications (scoring caps, penalty triggers and localized contests) on equilibrium behavior and output.

The equivalence noted by the last paper reveals a connection between these works and those papers that look at competitive Bayesian persuasion: see, e.g. \cite{albrecht}, \cite{cotton}, \cite{AU2}, and \cite{hkb}. Indeed for a binary prior and uncorrelated states, the competitive persuasion problem and the contest of \cite{fang} (with a performance cap) are equivalent.\footnote{Interestingly, despite their mathematical similarities, neither the competitive persuasion papers nor the risk-taking contest papers seem to be aware of the other. Moreover, both strands of the literature are apparently nescient of \cite{spiegler}, who looks at the mathematically analogous problem of firms pricing to boundedly rational consumers.} At the end of Section \ref{compstat}, we discuss our findings in the context of competitive persuasion.

Naturally, this paper is also related to the substantial collection of papers that look at risk-taking contests more broadly. This group of papers includes \cite{dasgupta1980uncertainty}, \cite{bhattacharya1986portfolio} and \cite{klette1986market}, who look at variants of an R\&D contest; \cite{hvide2002tournament}, \cite{hvide2003risk}, \cite{goel2008overconfidence}, \cite{gilpatric2009risk}, and \cite{fang2021less}, who look at promotion contests; \cite{basak2015competition}, \cite{strack2016risk}, \cite{whitmeyer2019relative}, and \cite{lacker2019mean}, who look at contests between investment managers in financial settings; and \cite{robson1992status}, \cite{hopkins2018inequality}, and \cite{zhang2020pre}, who look at contests for status and/or relative rank. 

\section{The Main Analysis}\label{main}

Formally, our model is as follows. There are $n$ agents $i = 1, 2, \dots, n$ who participate in the following contest. Time is continuous, and at each point in time $t \geq 0$ each agent $i$ privately observes the realization of a stochastic process
\[X_{t}^{i} = x_{0} + \mu t + \sigma B_{t} \text{ ,}\]
where $x_0 > 0$ is the initial value for each agent's process. At any point in time, an agent may (privately) decide whether to stop the process and either submit the realization (and incur a cost $c \geq 0$) or decline to submit. If the agent submits her output and it is strictly greater than the maximal submitted output of the other contestants, she obtains a prize that we normalize to $1$. If there is a tie for first place, it is broken fairly (though this will not happen on the equilibrium path). If she does not submit her output, she gets a payoff of $0$.\footnote{Our results do not change qualitatively if a contestant is still eligible for the prize when she does not submit (more on this at the end of Section \ref{compstat}).}

We assume that the principal commits \textit{ex ante} to the submission fee, and is limited to costs within some interval $\left[0,\bar{c}\right]$ with $\bar{c} \in \left(0,1\right)$. Moreover, following the literature on this problem, we assume that $0$ is an absorbing boundary. We also make the following parametric assumption, which ensures that agents' equilibrium stopping times are finite:
\[1+n\frac{\left(1-\bar{c}\right)\left(\exp{\left\{-\frac{2\mu x_0}{\sigma^2}\right\}}-1\right)}{1-n\bar{c} + \left(n-1\right)\bar{c}^{\left(\frac{n}{n-1}\right)}} > 0 \text{ .}\]
This is always satisfied when $\mu \leq 0$.

A strategy for an agent $i$ in the stopping problem is a stopping time $\tau_i$. Equivalently, because it is only the distribution over stopped values that matters, this problem of choosing $\tau_i$ can be reduced to one of choosing an optimal distribution $F_i$ over stopped values that is feasible, i.e., that can be induced by a stopping time. This problem of finding a stopping time to embed a probability measure is the well-known Skorokhod embedding problem.\footnote{\cite{obloj2004skorokhod} surveys the literature on this problem.} Moreover, the set of feasible distributions is readily available for us to take ``off-the-shelf.'' Let $\mathcal{F}$ be the set of feasible distributions and $s\left(x\right)$ be the scale function of the stochastic process $X$. For the Brownian motion with drift of this paper, the scale function is
\[\label{eq1}\tag{$1$}s_B\left(x\right) = \frac{\sigma^2}{2\mu}-\frac{\sigma^2}{2\mu}\exp{\left\{\frac{-2\mu x}{\sigma^{2}}\right\}} \text{ .}\]
Then, from Theorem 2.1 in \cite{pedersen2001azema} (see also Lemma 1 in \hyperlink{ss}{SS}, the discussion on p.25 of \cite{feng2015gambling} and Lemma 2.1 in \hyperlink{nz}{NZ}), 

\begin{remark}
The set of feasible distributions $\mathcal{F}$ consists of all distributions $F$ supported on $\left[0,\infty\right)$ that satisfy $\int_{\mathbb{R}_{+}}s\left(x\right)dF\left(x\right) = s\left(x_0\right)$.
\end{remark}
Note that when $X$ is a martingale, the set of feasible distributions becomes those distributions $F$ on $\left[0,\infty\right)$ that have mean $x_0$. Moreover, because the scale function is strictly monotone, as noted by \hyperlink{nz}{NZ} and \cite{feng2015gambling}, it is without loss of generality to solve for the equilibrium (and verify uniqueness) in the driftless case. 

This problem (without a submission fee) is solved in \cite{fang}, and it is easy to see that the fee only alters the strategic interaction slightly. Indeed, observe that agents' distributions over values that they do not disclose must be atomless and such that the payoff for each agent in disclosed values is an affine function of the value. Accordingly, for $c > 0$, each agent must both disclose and not disclose with positive probability--the former because the distribution over disclosed values must be atomless, and the latter because an agent always benefits by deviating and disclosing if nobody else does. 

It is also clear that if an agent does not disclose value $x'$ at equilibrium, then she does not disclose any values $x \leq x'$. Thus, because a concavification argument eliminates all other possibilities, the only value that an agent does not disclose at equilibrium is $0$--her equilibrium distribution must place a mass point on $0$. Moreover, she must be exactly indifferent between disclosing at $0$ or not, i.e.,
\[F^{n-1}\left(0\right) - c = 0 \text{ .}\]

Thus, the unique symmetric equilibrium when $X$ is a martingale is for each agent to choose the following distribution:
\[\Tilde{F}\left(x\right) = \left(c + \lambda \frac{x}{x_0}\right)^{\left(\frac{1}{n-1}\right)}, \quad \text{on} \quad \left[0,\left(1-c\right)\frac{x_0}{\lambda}\right] \text{ ,}\]\
where
\[\lambda = \lambda\left(c\right) \coloneqq \frac{1-nc + \left(n-1\right)c^{\left(\frac{n}{n-1}\right)}}{n} \text{ .}\]
The agent does not disclose if and only if she obtains realization $0$. Substituting in the scale function, we generate the equilibrium distribution when the diffusion has drift:
\[\label{eq2}\tag{$2$}F\left(x\right) = \left(c + \lambda\frac{s_B\left(x\right)}{s_B\left(x_{0}\right)}\right)^{\left(\frac{1}{n-1}\right)}, \quad \text{on} \quad \underbrace{\left[0,\frac{-\sigma^2}{2\mu}\log{\left\{1-\frac{2\mu}{\sigma^2}\frac{\left(1-c\right)s_B\left(x_{0}\right)}{\lambda}\right\}}\right]}_{\coloneqq \left[0, \bar{x}\left(c\right)\right]} \text{ .}\]
\begin{proposition}
There exists a unique symmetric equilibrium. Each agent chooses the distribution $F$ given in Expression \ref{eq2} and does not submit her realization if and only if it is $0$.
\end{proposition}
More generally, when $X$ is a non-negative time homogeneous diffusion process with a constant initial value and the parametric assumptions specific to the process are in place to ensure agents choose finite stopping times, the unique equilibrium distribution is for each agent to choose distribution 
\[\tag{$3$}\label{eq3}G\left(x\right) = \left(c + \lambda \frac{s\left(x\right)}{s\left(x_0\right)}\right)^{\left(\frac{1}{n-1}\right)}, \quad \text{on} \quad \underbrace{\left[0,\psi\left(\left(1-c\right)\frac{s\left(x_0\right)}{\lambda}\right)\right]}_{\eqqcolon \left[0,\bar{x}\left(c\right)\right]}\text{ ,}\]
where $s\left(x\right)$ is the scale function of $X$ and $\psi\left(\cdot\right) \coloneqq s^{-1}\left(\cdot\right)$. In particular, when $X$ is exponential Brownian motion, \[dX_t = \mu X_tdt + \sigma X_t dW_t \text{ ,}\]
with $2\mu < \sigma^2$, the scale function is $s\left(x\right) = x^{\kappa}/\kappa$, where $\kappa \coloneqq 1 - 2\mu/\sigma^2$, and the equilibrium distribution is
\[H\left(x\right) = \left(c+\lambda\left(\frac{x}{x_{0}}\right)^{\kappa}\right)^{\left(\frac{1}{n-1}\right)}, \quad \text{on} \quad \left[0,\left(\frac{\left(1-c\right)}{\lambda}\right)^{\frac{1}{\kappa}}x_{0}\right]\text{ .}\]

\section{Comparative Statics}\label{compstat}
In this part, we follow Section 3 of \hyperlink{nz}{NZ} closely--the statements and proofs of Theorem \ref{firstmaintheorem} and the two subsequent corollaries are taken from \hyperlink{nz}{NZ} with minimal changes. Naturally, the mechanism used by the principal to effect change is different, yet we show that raising submission fees affects equilibrium play in the same manner as making the prizes more unequal. Of course, the prize schedule in this model is already as inequitable as possible, yet submission fees still can benefit the principal significantly. The proofs for this section may be found in Appendix \ref{app}.

\begin{figure}[tbp]
\centering
\begin{subfigure}{.5\textwidth}
  \centering
  \includegraphics[scale=.14]{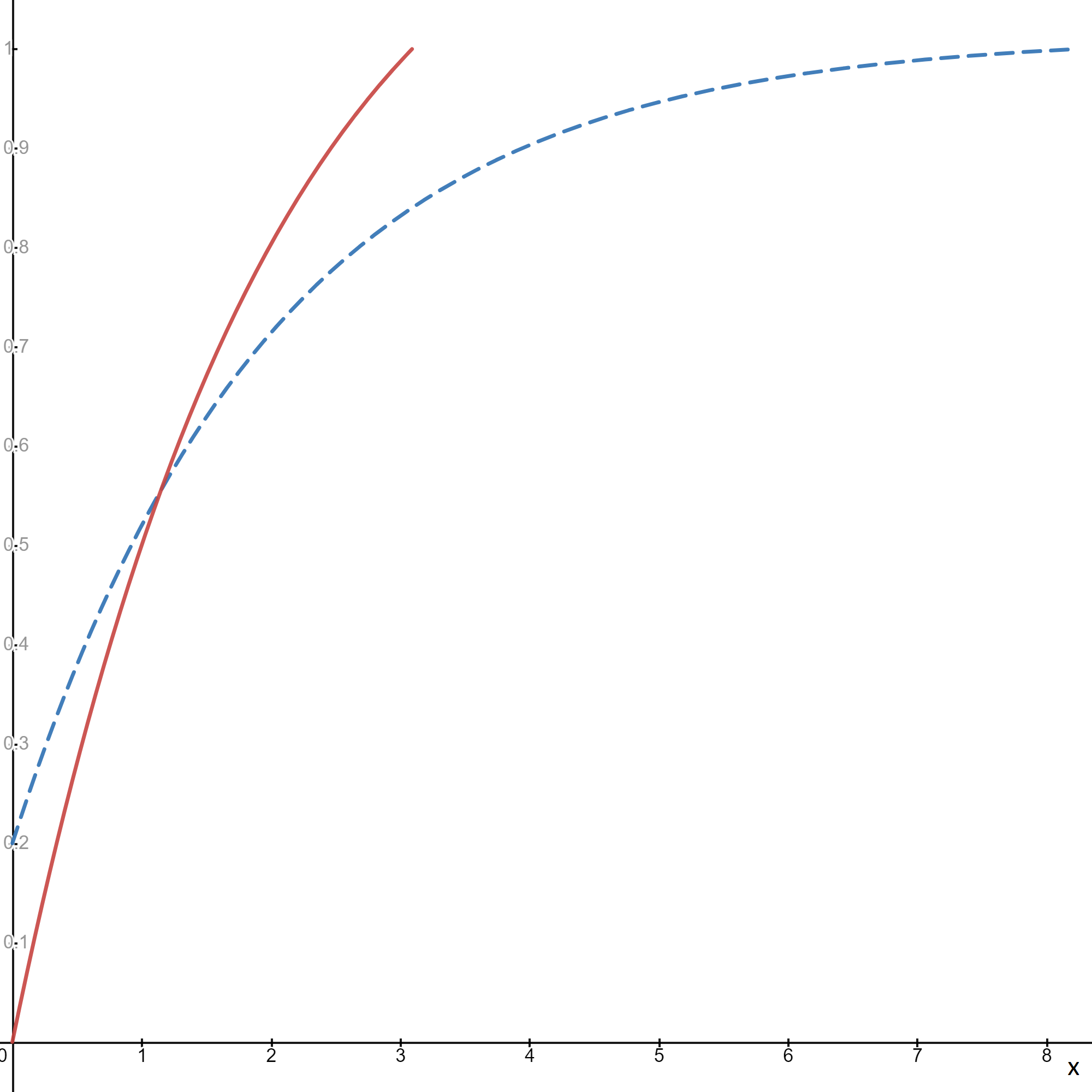}
  \caption{Brownian motion with drift.}
  \label{figsub1}
\end{subfigure}%
\begin{subfigure}{.5\textwidth}
  \centering
  \includegraphics[scale=.14]{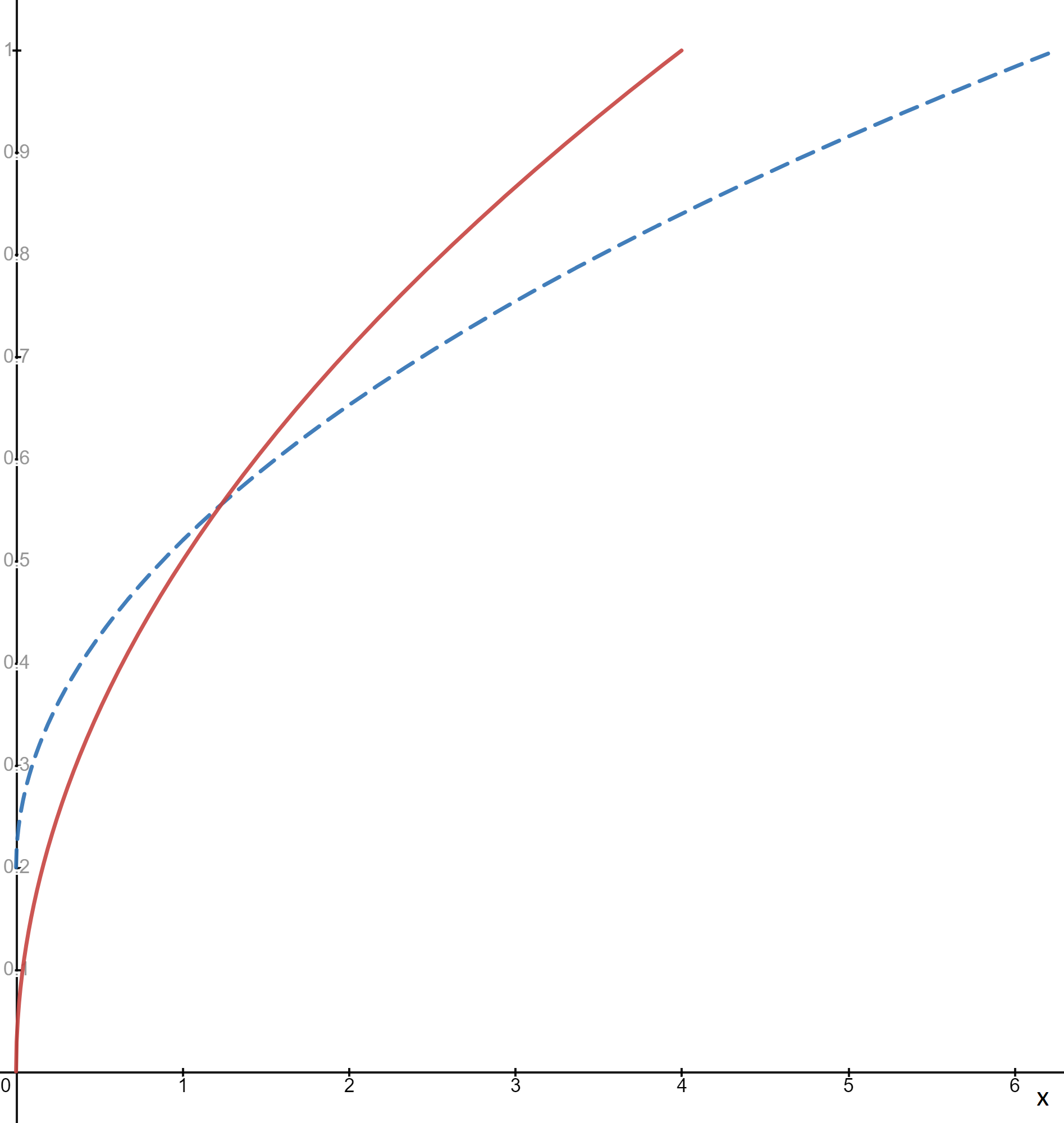}
  \caption{Exponential Brownian motion.}
  \label{figsub2}
\end{subfigure}
\caption{Equilibrium cdfs for Brownian motion with drift and exponential Brownian motion for $c = 0$ (solid red) and $c > 0$ (dashed blue).}
\label{fig1}
\end{figure}

We say that two cdfs $F$ and $\hat{F}$ are strictly single-crossing if there exists a point $x^{*} \in \supp{F}\cup\supp{\hat{F}}$ such that $F\left(x\right) > \hat{F}\left(x\right)$ for all $x \in \left[0,x^{*}\right)$ and $F\left(x\right) < \hat{F}\left(x\right)$ for all $x \in \left(x^{*},\bar{x}\left(c\right)\right)$ (with equality for all $x < 0$ and $x \geq \bar{x}\left(c\right)$ and at $x=x^{*}$).

\begin{lemma}\label{lemma23}
Let $c > \hat{c}$. Then the equilibrium distribution corresponding to $\hat{c}$, $\hat{F}$, is strictly single crossing with regard to the equilibrium distribution corresponding to $c$, $F$. 
\end{lemma}

This single crossing phenomenon can be seen in Figure \ref{fig1}. Using Corollary 2.5 of \cite{muller2017between}, Lemma \ref{lemma23} immediately implies the following result.

\begin{corollary}
Let $c > \hat{c}$ and set $\mu = 0$. Then the equilibrium distribution corresponding to $\hat{c}$, $\hat{F}$, second order stochastically dominates the equilibrium distribution corresponding to $c$, $F$. 
\end{corollary}

Next, we encounter the following theorem, which specifies the effect of submission fees on average equilibrium output.

\begin{theorem}\label{firstmaintheorem}
Let $c > \hat{c}$ and $F$ and $\hat{F}$ be the corresponding equilibrium distributions, respectively. Let $\phi \colon \mathbb{R}_{+} \to \mathbb{R}$ be an increasing, absolutely continuous function.
\begin{enumerate}
    \item If $\phi'/s'$ is increasing on $\left(0,\bar{x}\left(c\right)\right)$, then $\mathbb{E}_{\hat{F}}\left[\phi\left(x\right)\right] \leq \mathbb{E}_{F}\left[\phi\left(x\right)\right]$.
    \item If $\phi'/s'$ is decreasing on $\left(0,\bar{x}\left(c\right)\right)$, then $\mathbb{E}_{\hat{F}}\left[\phi\left(x\right)\right] \geq \mathbb{E}_{F}\left[\phi\left(x\right)\right]$.
\end{enumerate}
The inequalities are strict unless $\phi$ is an affine transformation of $s$.
\end{theorem}

For Brownian motion with drift, $s'\left(x\right) = \exp{\left\{-2\mu x/\sigma^2\right\}}$, and so $d/dx \left[\phi'\left(x\right)/s'\left(x\right)\right]$ has the same sign as $\phi''\left(x\right) + 2\mu/\sigma^2 \phi'\left(x\right)$. Similarly, for exponential Brownian motion, $s'\left(x\right) = x^{-2\mu/\sigma^2}$, and so 
$d/dx \left[\phi'\left(x\right)/s'\left(x\right)\right]$ has the same sign as $\phi''\left(x\right) + 2\mu/\sigma^2 \phi'\left(x\right)/x$. Thus,
\begin{corollary}\label{corr34}
When agents' processes are either Brownian motion with drift or exponential Brownian motion, $\mathbb{E}\left[\phi\left(x\right)\right]$ is increasing in the submission fee when $\mu \geq 0$ and $\phi$ is convex and decreasing in the submission fee when $\mu \leq 0$ and $\phi$ is concave. These inequalities are strict if $\phi$ is not constant and $\mu \neq 0$.
\end{corollary}

This immediately implies the following result concerning the effect of submission fees on the average output of each agent.

\begin{corollary}\label{corr27}
When agents' processes are either Brownian motion with drift or exponential Brownian motion, the average performance of an agent, $\mathbb{E}\left[X_{\tau}\right]$, is strictly increasing in the submission fee when $\mu > 0$ and strictly decreasing in the submission fee when $\mu < 0$. Thus, the uniquely optimal submission fees, respectively, are $\bar{c}$ and $0$.
\end{corollary}
One interesting relative of this result concerns the magnitude of the improvement in output that can be brought about via a submission fee. Namely, for certain regions of the parameter space the total output with two contestants and the optimal submission fee exceeds the total output with three contestants and no submission fee.

Our final result of this section concerns the expected value of the contest winner's performance. As one might suspect, the expected value of the maximal output is increasing in the size of the submission fee when the drift is positive. Intuitively, when the principal cares about the maximum from a number of draws from a distribution, she prefers longer right tails. When the drift is positive, she also prefers such longer experimentation and so the effects act in synergy. On the other hand, when the drift is negative, the two forces are possibly countervailing since increasing the second moment sacrifices the first. However, it turns out that the latter is more important: the principal always prefers a small submission fee to no submission fee.

\begin{theorem}\label{thm28}
When agents' processes are Brownian motion with drift or exponential Brownian motion, the submission fee that maximizes the expected maximal performance, $\mathbb{E}\left[\max_{i}X_{\tau_i}\right]$, is $\bar{c}$ if $\mu \geq 0$ and is strictly greater than $0$ if $\mu < 0$.
\end{theorem}

In this paper's model, the agents who do not submit entries are ineligible for the prize. It is easy to extend our analysis to the modified scenario in which agents who do not submit entries are still eligible (where their values are now their expected values, at equilibrium, given their choice of non-disclosure). Moreover, in some contests, submission fees are refunded in the event of a success; and in some contests, submission fees are refunded in the event of a failure. Those changes are also easy to accommodate.

In the first modification, in which agents who do not submit entries are still eligible, it is obvious that the belief assigned by the principal to non-disclosure must be $0$ (since otherwise an agent would benefit by secretly experimenting more in the non-disclosure region). As in the main specification, each agent must put a mass point on $0$ but now the size of the mass point $\gamma_{N}$ must solve
\[\gamma_N^{n-1} - c = \frac{\gamma_N^{n-1}}{n} \text{ ;}\]
\textit{viz.}, $\gamma_N^{n-1} = n c/\left(n-1\right)$. In the second modification, in which only submitters are eligible, but obtain a refund in the event of a victory, the size of the mass point $\gamma_{V}$ must solve
\[\gamma_{V}^{n-1}\left(1+c\right) - c = 0\text{ ;}\]
that is, $\gamma_{V}^{n-1} = c/\left(1+c\right)$. In the third modification, in which only submitters are eligible, but obtain a refund in the event of a loss, the size of the mass point $\gamma_{L}$ must solve
\[\gamma_{L}^{n-1} + (1 - \gamma_{L}^{n-1})c - c = 0\text{ ;}\]
which reduces to $\gamma_{L} = 0$. Given these, the unique equilibrium distribution in the three modifications when the process is Brownian motion with drift is
\[F\left(x\right) = \left(\gamma_{\iota} + \lambda_{\iota}\frac{s_B\left(x\right)}{s_B\left(x_{0}\right)}\right)^{\left(\frac{1}{n-1}\right)}, \quad \text{on} \quad \left[0,\frac{-\sigma^2}{2\mu}\log{\left\{1-\frac{2\mu}{\sigma^2}\frac{\left(1-\gamma_{\iota}\right)s_B\left(x_{0}\right)}{\lambda_{\iota}}\right\}}\right], \ \iota = N, V, L \text{ ,}\]
where 
\[\lambda_{\iota} \coloneqq \frac{1-n\gamma_\iota + \left(n-1\right)\gamma_\iota^{\left(\frac{n}{n-1}\right)}}{n}, \ \iota = N, V, L \text{ .}\]
In short, changing the particulars regarding the submission fee is equivalent to changing the cost. The analogous construction holds when the process is exponential Brownian motion. Finally, observe that for $c > 0$,
\[\underbrace{0}_{\gamma_{L}^{n-1}} < \underbrace{\frac{c}{1+c}}_{\gamma_{V}^{n-1}} < c < \underbrace{\frac{nc}{n-1}}_{\gamma_{N}^{n-1}} \text{ ,}\]
and so the following remark follows from the earlier results.
\begin{remark}
For a fixed $c$, when agents' processes are either Brownian motion with drift or exponential Brownian motion, the average performance of an agent, $\mathbb{E}\left[X_{\tau}\right]$, is highest when non-submitters are also eligible for the prize when $\mu \geq 0$ and is highest when losers' fees are refunded when $\mu < 0$.
\end{remark}
Interestingly, it is never optimal to refund the winner's submission fee but it may be optimal to refund the submission fees of the losers, though this is preferred if and only if no submission fee is best.

It is immediate that the special case of the contest when the drift equals zero and non-submitters are eligible for the prize is equivalent to a competitive persuasion problem in which $n$ agents each privately choose experiments about an idiosyncratic binary state, which they may disclose to a receiver at some cost $c \geq 0$, who then selects the sender whom he esteems highest, breaking ties fairly. The lone modification is that, because agents' values are beliefs, there is an upper bound of $1$ on agent's values, so for large enough $n$ or prior value, $x_0$, agents place a mass point on value $1$ as well. Thus, an analog of Theorem \ref{thm28} holds: for all disclosure costs sufficiently small, the principal's (receiver's) welfare is strictly increasing in agents' disclosure costs.

\section{Discussion}

This paper establishes that in contests in which agents compete by randomization, introducing submission fees benefits a principal whenever she prefers more experimentation. Thus, in a variety of contests--including those pertaining to innovation, promotion within a firm, investment, status, and persuasion--seemingly inefficient frictions may actually improve a principal's (society's) welfare. Unlike entry fees, which may benefit a principal by increasing effort by dampening competition, submission fees do the opposite and fan competition's flames, encouraging greater risk taking and longer tails in agents' outputs.

\bibliography{sample.bib}

\appendix

\section{Omitted Proofs}\label{app}

\subsection{Lemma \ref{lemma23} Proof}

\begin{proof}
We prove this result for the general case in which agents' processes are non-negative time homogeneous diffusions. The equilibrium distribution $G$ is given in Expression \ref{eq3}.

Directly, $G\left(0\right) = c^{\left(\frac{1}{n-1}\right)}$, which is obviously strictly increasing in $c$. Likewise,
\[\frac{\partial{}}{\partial{c}}\left\{\psi\left(\left(1-c\right)\frac{s\left(x_0\right)}{\lambda}\right)\right\} = \psi'\left(\left(1-c\right)\frac{s\left(x_0\right)}{\lambda}\right)s\left(x_0\right)\frac{\partial{}}{\partial{c}}\left\{\frac{1-c}{\lambda}\right\} > 0  \text{ ,}\]
since $s$ is strictly monotone. As a result, by the intermediate value theorem, there exists at least one point $x$ in the interior of $\supp{G}\cup\supp{\hat{G}}$ where $G\left(x\right) = \hat{G}\left(x\right)$.

Next, we need to show that this intersection point is unique. Define $\Upsilon\left(x\right) \coloneqq G\left(x\right) - \hat{G}\left(x\right)$, which can be written out as
\[\Upsilon\left(x\right) = \left(c + \lambda \frac{s\left(x\right)}{s\left(x_0\right)}\right)^{\left(\frac{1}{n-1}\right)} - \left(\hat{c} + \hat{\lambda} \frac{s\left(x\right)}{s\left(x_0\right)}\right)^{\left(\frac{1}{n-1}\right)} \text{ ,}\]
where $\hat{\lambda}$ is defined in the obvious way. Directly, $\Upsilon'\left(x\right)$ has the same sign as 
\[\underbrace{\lambda\left(c + \lambda \frac{s\left(x\right)}{s\left(x_0\right)}\right)^{\left(\frac{1}{n-1} - 1\right)}}_{\eqqcolon r\left(\lambda\right)} - \hat{\lambda}\left(\hat{c} + \hat{\lambda} \frac{s\left(x\right)}{s\left(x_0\right)}\right)^{\left(\frac{1}{n-1} - 1\right)} \text{ .}\]
If $n=2$, this reduces to $\lambda - \hat{\lambda} < 0$, as required. For the remainder let $n \geq 3$. Then,
\[\begin{split}
    r'\left(\lambda\right) &= \left(c + \lambda \frac{s\left(x\right)}{s\left(x_0\right)}\right)^{\left(\frac{1}{n-1} - 1\right)} + \lambda \frac{s\left(x\right)}{s\left(x_0\right)}\left(\frac{1}{n-1}-1\right)\left(c + \lambda \frac{s\left(x\right)}{s\left(x_0\right)}\right)^{\left(\frac{1}{n-1} - 2\right)}\\
    &= \left(c + \lambda \frac{s\left(x\right)}{s\left(x_0\right)}\frac{1}{n-1}\right)\left(c + \lambda \frac{s\left(x\right)}{s\left(x_0\right)}\right)^{\left(\frac{1}{n-1} - 2\right)} > 0
\end{split} \text{ .}\]
Now, suppose for the sake of contradiction that $\Upsilon'\left(x\right) \geq 0$, which implies that 
\[\hat{\lambda}\left(c + \hat{\lambda} \frac{s\left(x\right)}{s\left(x_0\right)}\right)^{\left(\frac{1}{n-1} - 1\right)} - \hat{\lambda}\left(\hat{c} + \hat{\lambda} \frac{s\left(x\right)}{s\left(x_0\right)}\right)^{\left(\frac{1}{n-1} - 1\right)} \geq 0 \text{ ,}\]
or
\[\left(c + \hat{\lambda} \frac{s\left(x\right)}{s\left(x_0\right)}\right)^{-\left(\frac{n-2}{n-1}\right)} \geq \left(
\hat{c} + \hat{\lambda} \frac{s\left(x\right)}{s\left(x_0\right)}\right)^{-\left(\frac{n-2}{n-1}\right)} \text{ ,}\]
a contradiction because $c > \hat{c}$.
\end{proof}

\subsection{Theorem \ref{firstmaintheorem} Proof}
\begin{proof}
As we do for the previous lemma, we establish this result for the general case. The equilibrium distribution $G$ is that described in Expression \ref{eq3}. We follow the proof of Theorem 3.2 in \hyperlink{nz}{NZ} virtually verbatim.

Via integration by parts,
\[\mathbb{E}_{\hat{G}}\left[\phi\left(x\right)\right] - \mathbb{E}_{G}\left[\phi\left(x\right)\right] = -\int_{0}^{\bar{x}\left(c\right)}\left(\hat{G}\left(x\right) - G\left(x\right)\right)\phi'\left(x\right)dx \text{ .}\]
By Lemma \ref{lemma23}, $\hat{G}$ is strictly single-crossing with regard to $G$ with some intersection point $x^{*} \in \left(0,\bar{x}\left(c\right)\right)$. Thus,
\[\int_{0}^{\bar{x}\left(c\right)}\left(\hat{G}\left(x\right) - G\left(x\right)\right)s'\left(x\right)\frac{\phi'\left(x\right)}{s'\left(x\right)}dx \geq \frac{\phi'\left(x^{*}\right)}{s'\left(x^{*}\right)}\int_{0}^{\bar{x}\left(c\right)}\left(\hat{G}\left(x\right) - G\left(x\right)\right)s'\left(x\right)dx \text{ .}\]
Next, we again integrate by parts
\[\int_{0}^{\bar{x}\left(c\right)}\left(\hat{G}\left(x\right) - G\left(x\right)\right)s'\left(x\right)dx = \left.\left(\hat{G}\left(x\right) - G\left(x\right)\right)s\left(x\right)\right\vert_{0}^{\bar{x}\left(c\right)} - \int_{0}^{\bar{x}\left(c\right)}\left(\hat{g}\left(x\right) - g\left(x\right)\right)s\left(x\right)dx= 0 \text{ ,}\]
since by definition the feasible distributions, $G$, are those that satisfy 
\[\int_{\mathbb{R}_{+}}\frac{s\left(x\right)}{s\left(x_{0}\right)}dG\left(x\right) = 1 \text{ .}\]
Combining expressions yields part 1 of the theorem, and the second part can be obtained in virtually identical fashion \textit{mutatis mutandis}.
\end{proof}

\subsection{Theorem \ref{thm28} Proof}

\begin{proof}
We keep the convention that $c > \hat{c} \geq 0$ with hats over the corresponding objects to the latter. By Lemma \ref{lemma23} $\hat{F}$ is strictly single crossing with regard to $F$ and so $F^{-1}$ is strictly single crossing with regard to $\hat{F}^{-1}$. Let $y_{*}$ be the unique crossing point. First, let $\mu \geq 0$, in which case we may follow the proof of Theorem 3.8 in \hyperlink{nz}{NZ}. For Brownian motion with drift or exponential Brownian motion we have
\[\int_{c^{\left(\frac{1}{n-1}\right)}}^{1}y^{n-1}F^{-1}\left(y\right) dy - \int_{\hat{c}^{\left(\frac{1}{n-1}\right)}}^{1}y^{n-1}\hat{F}^{-1}\left(y\right) dy > y_{*}^{n-1}\left[\int_{c^{\left(\frac{1}{n-1}\right)}}^{1}F^{-1}\left(y\right) dy - \int_{\hat{c}^{\left(\frac{1}{n-1}\right)}}^{1}\hat{F}^{-1}\left(y\right) dy\right] \geq 0 \text{ ,}\]
where the last inequality follows from Corollary \ref{corr27} since $\mu \geq 0$. 

Now let $\mu < 0$, and observe that it suffices to show that $\mathbb{E}\left[\max_i{X_i}\right]$ is strictly increasing in $c$ at $c = 0$. Via Leibniz's rule, we have
\[\label{eqa1}\tag{$A1$}\frac{d}{dc}\left\{\int_{c^{\left(\frac{1}{n-1}\right)}}^{1}y^{n-1}F^{-1}\left(y\right) dy\right\} = \int_{c^{\left(\frac{1}{n-1}\right)}}^{1}\frac{\partial}{\partial{c}}\left\{y^{n-1}F^{-1}\left(y\right)\right\} dy \text{ .}\]
When the process is exponential Brownian motion
\[F^{-1}\left(y\right) = x_0 \left(\frac{y^{n-1} - c}{\lambda}\right)^{\left(\frac{1}{\kappa}\right)} \text{ ,}\]
where, recall, $\kappa = 1 - 2\mu/\sigma^2$. Using this, when $\mu < 0$ Expression \ref{eqa1} evaluated at $c = 0$ has the same sign as
\[\int_{0}^{1}y^{n-1}\left(y^{n-1}\right)^{\left(\frac{1}{\kappa} - 1\right)}\left(-1 + n y^{n-1}\right)dy = \frac{\kappa\left(n-1\right)^2}{\left(n+\kappa-1\right)\left(\kappa n+n-1\right)} > 0 \text{ .}\]

When the process is Brownian motion with drift,
\[F^{-1}\left(y\right) = -\alpha \ln{\left\{1-\frac{\left(y^{n-1}-c\right)s\left(x_0\right)}{\lambda \alpha}\right\}} \text{ ,}\]
where $\alpha \coloneqq \sigma^2/\left(2\mu\right)$. Using this, Expression \ref{eqa1} evaluated at $0$ has the same sign as
\[\int_{0}^{1}\frac{n y^{n-1} - 1}{1 + n \beta y^{n-1}} y^{n-1}dy > \frac{\frac{1}{n}}{1 + \beta}\int_{0}^{1}\left(n y^{n-1} - 1\right)dy = 0 \text{ ,}\]
where $\beta \coloneqq -\left(1-\exp{\left\{-x_0/\alpha\right\}} \right) > 0$. 
\end{proof}

\end{document}